\documentclass[a4paper,11pt]{article}
\pdfoutput=1 % if your are submitting a pdflatex (i.e. if you have
\usepackage{jcappub} % for details on the use of the package, please
\usepackage[T1]{fontenc} % if needed
\usepackage{amsmath, amssymb, bm, graphicx, graphics, color, mathrsfs}
\RequirePackage{graphicx}
\RequirePackage{mathptmx}      % use Times fonts if available on your TeX system
\RequirePackage{flushend}
\RequirePackage[numbers,sort&compress]{natbib}

\title{\boldmath Primordial density fluctuations generated by wormholes }

\author[1,a]{Marek Rogatko\note{rogat@kft.umcs.lublin.pl, marek.rogatko@poczta.umcs.lublin.pl}}
\author[2,a,b]{Aneta Wojnar\note{aneta.wojnar@cosmo-ufes.org}}
\author[3,a]{Bart{\l}omiej Kiczek\note{bkiczek@kft.umcs.lublin.pl}}
\affiliation[a]{Institute of Physics \\
Maria Curie-Sk{\l}odowska University \\
20-031 Lublin, pl. Marii Curie-Sk{\l}odowskiej 1, Poland}
\affiliation[b]{N{\'u}cleo Cosmo-ufes \& PPGCosmo, Universidade Federal do Esp{\'i}rito Santo,
29075-910, Vit{\'o}ria, ES, Brasil.}
\abstract{
We studied the influence of a wormhole existing in the expanding Universe and passing through the horizon volume
on the energy density of the Universe. The two-point correlation function of a free massless scalar field mimicking the inflaton was found.
The revealed violation of the translation invariance during the process in question might be potentially  envisaged in the anisotropy in cosmic 
microwave background radiation map.

}

\begin{document}
\maketitle
\flushbottom
%%%%%
\newcommand  {\Rbar} {{\mbox{\rm$\mbox{I}\+!\mbox{R}$}}}
\newcommand  {\Hbar} {{\mbox{\rm$\mbox{I}\!\mbox{H}$}}}
\newcommand {\Cbar}{\mathord{\setlength{\unitlength}{1em}
     \begin{picture}(0.6,0.7)(-0.1,0) \put(-0.1,0){\rm C}
        \thicklines \put(0.2,0.05){\line(0,1){0.55}}\end {picture}}}
%%%%%%%%%%%%%%%%%%%%%%%%%%%%%%%%%%%%%%%%%%%%%%%%%%%%%%%%%%%%%%%%%%%%%
% other new commands
\newcommand{\be}{\begin{equation}}
\newcommand{\ee}{\end{equation}}
\newcommand{\ben}{\begin{eqnarray}}
\newcommand{\een}{\end{eqnarray}}

\newcommand{\la}{{\lambda}}
\newcommand{\Om}{{\Omega}}
\newcommand{\ta}{{\tilde a}}
\newcommand{\bg}{{\bar g}}
\newcommand{\bh}{{\bar h}}
\newcommand{\bdel}{{\bar \delta}}
\newcommand{\si}{{\sigma}}
\newcommand{\C}{{\cal C}}
\newcommand{\D}{{\cal D}}
\newcommand{\cA}{{\cal A}}
\newcommand{\cT}{{\cal T}}
\newcommand{\cO}{{\cal O}}
\newcommand{\eeo}{\cO ({1 \over E})}
\newcommand{\G}{{\cal G}}
\newcommand{\cL}{{\cal L}}
\newcommand{\cH}{{\cal H}}
\newcommand{\cE}{{\cal E}}
\newcommand{\cM}{{\cal M}}
\newcommand{\cR}{{\cal R}}
\newcommand{\cB}{{\cal B}}
\newcommand{\cJ}{{\cal J}}

\newcommand{\p}{\partial}
\newcommand{\na}{\nabla}
\newcommand{\ssum}{\sum\limits_{i = 1}^3}
\newcommand{\dssum}{\sum\limits_{i = 1}^2}
\newcommand{\tal}{{\tilde \alpha}}
\newcommand{\ints}{\int_{\Sigma} d\Sigma}
\newcommand{\LieN}{{\cal L}_{N^{i}}}
\newcommand{\Lief}{{\cal L}_{\phi^{i}}}
\newcommand{\Liet}{{\cal L}_{t^{i}}}
\newcommand{\LieM}{{\cal L}_{M^{\mu}}}
\newcommand{\Lie}{{\cal L}}

\newcommand{\tpe}{{\tilde p}}
\newcommand{\tp}{{\tilde \phi}}
\newcommand{\tPhi}{\tilde \Phi}
\newcommand{\tpsi}{\tilde \psi}
\newcommand{\tchi}{\tilde \chi}
\newcommand{\tim}{{\tilde \mu}}
\newcommand{\tom}{{\tilde \omega}}
\newcommand{\tr}{{\tilde \rho}}
\newcommand{\tV}{{\tilde V}}
\newcommand{\tir}{{\tilde r}}
\newcommand{\rp}{r_{+}}
\newcommand{\hr}{{\hat r}}
\newcommand{\rv}{{r_{v}}}
\newcommand{\dr}{{d \over d \hr}}
\newcommand{\dR}{{d \over d R}}

\newcommand{\hhf}{{\hat \phi}}
\newcommand{\hhM}{{\hat M}}
\newcommand{\hhQ}{{\hat Q}}
\newcommand{\hht}{{\hat t}}
\newcommand{\hhr}{{\hat r}}
\newcommand{\hhS}{{\hat \Sigma}}
\newcommand{\hhD}{{\hat \Delta}}
\newcommand{\hhm}{{\hat \mu}}
\newcommand{\hro}{{\hat \rho}}
\newcommand{\hhz}{{\hat z}}

\newcommand{\hI}{\hat I}
\newcommand{\hg}{\hat g}
\newcommand{\hR}{\hat R}
\newcommand{\hD}{\hat D}
\newcommand{\hna}{\hat \nabla}
\newcommand{\tF}{\tilde F}
\newcommand{\tT}{\tilde T}
\newcommand{\tL}{\tilde L}
\newcommand{\hC}{\hat C}

\newcommand{\ep}{\epsilon}
\newcommand{\tep}{\tilde \epsilon}
\newcommand{\bep}{\bar \epsilon}
\newcommand{\ppp}{\varphi}
\newcommand{\Ga}{\Gamma}
\newcommand{\ga}{\gamma}
\newcommand{\hth}{\hat \theta}
\newcommand{\zpsi}{\psi^{\ast}}

\newcommand{\Dsl}{{\slash \negthinspace \negthinspace \negthinspace \negthinspace  D}}
\newcommand{\tD}{{\tilde D}}
\newcommand{\tB}{{\tilde B}}
\newcommand{\talpha}{{\tilde \alpha}}
\newcommand{\tbeta}{{\tilde \beta}}
\newcommand{\hT}{\hat T}

%%%%%%%%%%%%%%%%%%%%%%%%%%%%%%%%%%%%%%%%%%%%%%%%%%%%%%%%%%%%%%%%%%%%%%%%
%%%%%%%%%%%%%%%%%%%%%%%%%%%%%%%%%%%%%%%%%%%%%%%%%%%%%%%%%%%%%%%%%%%%%%%%%%%

\section{Introduction}
\label{sec:intro}
Contemporary cosmology consists of a few hypothesis and paradigms. The crucial one is the cosmological principle claiming that 
the Universe is approximately homogeneous and isotropic on large scales, and assuming the matter and geometry of spacetime are 
represented by smooth function of a position.
On the other hand, our Universe is fulfilled with the cosmic microwave background (CMB) radiation, as observed today, almost homogenous and isotropic.
The tiny fluctuations of CMB radiation are of order $10^{-5}$ \cite{smoot}.
Those deviations of CMB radiation appeared after the recombination epoch of our Universe when the photons decoupled from the electrons
making the Universe transparent. It constitutes radiation of free photons with a perfect blackbody spectrum \cite{ruth} which has been very well verified 
observationally \cite{sta18}. 
Measurements of the anisotropies in CMB radiation provide us information about the cosmological model describing the contents and evolution of the Universe.
The data delivered by them led to a standard cosmological model with only six parameters. The key CMB observables like temperature, polarization
anisotropies play the main role in the recent understanding of the initial conditions and structure formations inside the Universe.
The fluctuations on the large scale seen in the CMB spectrum allow us to determine the deviations from the  
Friedmann-Robertson-Lemaitre-Walker (FRLW) geometry. 

The question arises (the so-called horizon problem) why we observe the same background temperature in regions not being in causal contact before the CMB photons could freely
move through the spacetime. The evolution of the Universe provided by the $\Lambda$CDM model does not give enough time for creating so similar 
patches with the same temperature. Another ones are the flatness problem and monopole problem (unwanted relics) - the latter is 
predicted by models of particle physics giving that at very high temperatures there appear some artifacts which are not visible in our Universe presently. 

The inflationary cosmology, being the standard paradigm, proposed as a solution to the horizon, flatness and monopole problems \cite{starob, guth},
delivers a very generous mechanism for generating primordial perturbations, which result in large-scale structures and temperatures
anisotropies of CMB radiation. The observed departures from cosmological homogeneity and isotropy observed in CMB radiation
are so tiny that one may conclude that they should be dominated by a Gaussian probability distribution.

The inflationary phase, characterized by the very rapid Universe expansion, is assumed to be dominated by a scalar field (the most popular is given by the so-called inflaton field \cite{LL}), which quantum fluctuations provide 
density fluctuations reproducing the almost scale invariant spectrum of classical density perturbations \cite{komatsu}. 

The cosmic of no-hair theorem \cite{gib77} claims that the inflationary Universe asymptotically tends to the de-Sitter spacetime till the end of the inflationary phase.
However, it was recently claimed that the effects of matter and spacetime inhomogeneities to inflation should be taken into account \cite{cho}.

%%%%%%%%%%%%%%%%%%%%%%%%%%%%%%%%%%%%%%%%%%%%%%%%%%%%%%%%%%%%%%%%%%%%%%%%%%%%

 In the simplest form one deals with the definition of 
power spectrum in Fourier space of the primordial density perturbation $\delta(\mathbf{k})$ in the following form\footnote{Later on we will assume
that the density perturbations $\delta(\mathbf{k})=\kappa\chi$.}
\begin{equation}
 \langle\delta(\mathbf{k})\delta(\mathbf{q})\rangle=P(k)(2\pi)^3\delta(\mathbf{k+q}),
\label{dd}
\end{equation}
or
\begin{equation}
 \langle\delta(\mathbf{x})\delta(\mathbf{y})\rangle=\int d^3k e^{i\mathbf{k\cdot(x-y)}} P(k),
\end{equation}
where we set $k=|\mathbf{k}|$ and $P$ is the power spectrum. 

The form of the relation (\ref{dd}) is connected with rotational and translational invariance of the two-point correlation of the Fourier transform taken
from the primordial density $\delta( \mathbf{k})$. Namely,
the dependence of $P$ on the magnitude of the wave-vector $k$ informs
about the rotational invariance, while the delta function presence of $\mathbf{k+q}$ is a consequence of the translational 
invariance. Now on, when we suppose that there could exist some objects during the inflationary era which violated the invariance, they should leave imprints on the
anisotropy of the CMB radiation caused by such a violation. Therefore, it was suggested \cite{ack} how to compute the power spectrum of fluctuations 
in the case of a small violation of statistical isotropy which is characterized by a preferred direction in space along a unit vector $\mathbf{n}$. Thus, the primordial
power spectrum which takes into account the effects of the rotational invariance violation is now expressed (a parity $\mathbf{k\rightarrow-k}$ is assumed)
\begin{equation}
 P'(\mathbf{k})=P(k)\Big(1+g(\mathbf{\hat{k}\cdot n})^2\Big),
\end{equation}
where $\mathbf{\hat{k}}$ is the unit vector along the direction of $\mathbf{k}$ and $g$ is a $k$-independent constant. Further, the similar examination was done for a small 
violation of translational invariance \cite{carrol}. Similarly, they assume that an object breaking the invariance during inflation disappears after the end of the 
inflationary phase. They obtained expressions for the translational invariance broken by a special point, line or plane with the focus on the first one. The general form of 
the density perturbation's two-point correlation breaking statistical translational invariance by a presence of a special point $\mathbf{x}_0$ preserving rotational 
invariance about that point is provided by
\begin{equation}
 \langle\delta(\mathbf{x})\delta(\mathbf{y})\rangle=\int d^3k\int d^3q e^{i\mathbf{k\cdot(x-x_0)}} e^{i\mathbf{q\cdot(y-x_0)}} P_t(k,q,\mathbf{k\cdot q}),
\end{equation}
where $P_t$ is symmetric under interchange of $k$ and $q$. Consistent with the data, these violations are small resulting that $P_t$ is strongly peaked 
about $\mathbf{k=-q}$. A special line, which one identifies via a point $\mathbf{x}_0$ and a unit tangent vector $\mathbf{n}$
(for simplicity one locates it along the $z$ axis) preserves the rotation invariance about the distinguished axis, while the translational invariance along this preferred 
direction is left broken
\begin{equation}
  \langle\delta(\mathbf{x})\delta(\mathbf{y})\rangle=\int dk_z \int d k_\perp \int d^2q_\perp e^{ik_z(x_z-y_z)}e^{i\mathbf{k_\perp(\cdot x_\perp-x_{0\perp})}} 
  e^{i\mathbf{q_\perp\cdot (y_\perp-z_{0\perp}}}P_t(k_\perp,q_\perp,k_z,\mathbf{k_\perp\cdot q_\perp}),
\end{equation}
where $P_t$ is symmetric under interchange of $k_\perp$ and $q_\perp$. The similar expression can be obtained for the plane \cite{carrol}. Those relations enable us
to compute directly the correlations between spherical harmonic coefficients of the CMB temperature field, which can be used for the comparison the result of the 
CMB observations. 
%\textcolor{red}{tu moge przepisac wzory z Durrer, str. 95}

There are many candidates as the sources of anomalies, delivering small violation of rotational and translational invariance in the primordial
density fluctuations, which can be spotted in the CMB radiation pattern.
The impact of the preferred direction during the inflationary idea on the CMB anisotropy were studied in \cite{gum, arm, pereira, gum2, yoko, kan, wat}.
Quantum fluctuations of a free scalar field in the inflationary epoch were already investigated in a different context in the literature.
The point defect (Schwarzschild-de
Sitter black hole) were considered in \cite{cho}, while an infinitely long straight cosmic string was examined in \cite{string}. 

The main objective of the work is to shed some light on the problem of a small violation of translational and rotational invariances during the early phase of
our Universe, which can be in principle imprinted on the deviation from the statistical isotropy of the cosmic microwave background radiation.

The paper is organized as follows: in the section \ref{sec2} we derive the two-point correlation function of the massless scalar
field in the expanding universe with a wormhole. The numerical results and our conclusions are given in the section \ref{sec3}.

%%%%%%%%%%%%%%%%%%%%%%%%%%%%%%%%%%%%%%%%%%%%%%%%%%%%%%%%%%%%%%%%%%%%%%%%
%%%%%%%%%%%%%%%%%%%%%%%%%%%%%%%%%%%%%%%%%%%%%%%%%%%%%%%%%%%%%%%%%%%%%%%%

\section{The two-point correlation function of a massless scalar field}\label{sec2}
The primordial quantum fluctuations occurring during the inflation phase are statistically homogeneous, being the consequence of translation
invariance. However, the presence of compact objects, like cosmological defects, black holes or wormholes can potentially
lead to the violation of the aforementioned pattern, which produces the imprint on CMB radiation, leading to its anisotropy.

Our main aim in this section will be to find the two-point correlation function of scalar massless field which mimics the dilaton field during the inflation epoch.
 We shall be interested in the spacetime of a wormhole in the expanding universe whose existence one supposes in the inflationary era of the Universe
evolution. In our calculations we use the so-called $\langle in - in \rangle$ formalism \cite{Weinberg}, in which the Hamiltonian of the system, in the
interactive picture, will be found in a perturbative way. The unperturbed spacetime will authorize the de Sitter manifold.

The line element describing the object in question is provided by \cite{mir06}
\begin{equation}\label{metric}
 ds^2=-dt^2+a^2(t)\left(1+\frac{b^m(r)}{r^m}\right)^\frac{4}{m}(dr^2+r^2d\theta+dz^2),
\end{equation}
where by $a(t)=e^{Ht}$ an usual scale factor is denoted, while
$H$ stands for the Hubble parameter. 
The function 
$b(r)$ is the so-called shape function, as it determines the spatial shape of the wormhole. 
The $r$-coordinate decreases from infinity to the value $b_0$, being the size of the wormhole throat. 
Then it increases from it back to the infinity (to another universe or 
to some very distant place from ourselves).
Here, we are interested in the simplest wormhole, that is, when $b(r)=b=\text{const}$ and $m=1$. For 
convenience we denote $\alpha=\left(1+\frac{b}{r}\right)$. 

The other main ingredient in our consideration is massless scalar field which will mimic the inflaton field. Its action is given by the relation
\begin{equation}
S =- \frac{1}{2}\int d^4x\sqrt{-g} \na_\mu\chi \na^\mu\chi.                                                                    
\end{equation}
In the spacetime of interests the Lagrangian density of the dilaton field implies
\begin{equation}
 \mathcal{L}=\frac{ra^3\alpha^6}{2}\left(\dot{\chi}^2-\frac{(\partial_r\chi)^2}{\alpha^4 a^2}-\frac{(\partial_\theta\chi)^2}{\alpha^{4}a^{2}r^{2}}-
 \frac{(\partial_z\chi)^2}{a^{2}\alpha^{4}}
 \right).
\end{equation}
On the other hand, the Legendre transformation enables one to find the Hamiltonian of the system
\begin{equation}
 \mathcal{H}=\int d^3x~
 \frac{ra^2\alpha^6}{2}\left(\dot{\chi}^2+\frac{(\partial_r\chi)^2}{\alpha^4 a^2}+\frac{(\partial_\theta\chi)^2}{\alpha^{4}a^{2}r^{2}}+
 \frac{(\partial_z\chi)^2}{a^{2}\alpha^{4}}
 \right).
\end{equation}
In order to implement the $\langle in - in \rangle$ formalism we expand in a Taylor series the Hamiltonian, near
$b \rightarrow 0$. The result can be written as
\begin{equation}
 \mathcal{H}=\int rdrd\theta dz~
 \frac{a^3r^2\dot{\chi}^2+a(\partial_\theta\chi)^2+ar^2(\partial_r\chi)^2+ar^2(\partial_z\chi)^2}{2r^2}+\mathcal{H}_\text{int},
\end{equation}
where the first term is an unperturbed Hamiltonian in the de Sitter spacetime, while 
\begin{equation}\label{hami}
  \mathcal{H}_\text{int}=b\int d^3x \left( 3a^3\dot{\chi}^2+a(\partial_r\chi)^2+\frac{a}{r^2}(\partial_\theta\chi)^2+a(\partial_z\chi)^2
 \right).
\end{equation}
describes the interaction-picture Hamiltonian $\mathcal{H}_\text{int}(t)$, up to the first order in $b$ since in that case we have that 
$[\mathcal{H}_0,\mathcal{H}_\text{int}]=0$.

Upon the quantization the inflaton field $\chi$ becomes a quantum operator
\begin{equation}
 \chi_I(\mathbf{x},\tau)=\int\frac{d^3k}{(2\pi)^3}e^{ik_zz+ik_\perp r\cos(\theta-\theta_k)}\Big( \chi_{\mathbf{k}}(\tau)\beta(\mathbf{k}) +\chi^*_{\mathbf{k}}(\tau)\beta^\dag(\mathbf{-k}) \Big)
\end{equation}
where $\tau=-\frac{1}{{H}}e^{-{H} t}$ is a conformal time.
The conformal time varies from $- \infty$ to $0$, while $t$-coordinate changes from $-\infty$ to $\infty$.
$\beta(\mathbf{k})$ annihilates the vacuum state satisfying the standard commutation 
relations $[\beta(\mathbf{k}),\beta^\dag(\mathbf{-q})]=(2\pi)^3\delta(\mathbf{k}-\mathbf{q})$. Therefore, we rewrite the Hamiltonian (\ref{hami}) as
\begin{align}\label{int} \nonumber
 \mathcal{H}_\text{int}(\tau')&=b\int d^3x \frac{d^3k}{(2\pi)^3}\frac{d^3q}{(2\pi)^3}e^{i(\mathbf{k}+\mathbf{q})\mathbf{x}} \\ \nonumber
&{} \left\{ 3a^3(\tau')\Big( \dot{\chi}_{\mathbf{k}}(\tau)\beta(\mathbf{k}) +\dot{\chi}^*_{\mathbf{k}}(\tau)\beta^\dag(\mathbf{-k}) \Big)
\Big( \dot{\chi}_{\mathbf{q}}(\tau)\beta(\mathbf{q}) +\dot{\chi}^*_{\mathbf{q}}(\tau)\beta^\dag(\mathbf{-q}) \Big)\right.\nonumber\\ 
  &\left.+\left(-a(\tau')k_\perp q_\perp\cos(\theta-\theta_k)\cos(\theta-\theta_q)-a(\tau')k_zq_z-(k_xy-k_yx)(q_xy-q_yx)\frac{a(\tau')}{x^2+y^2}\right)\right.\nonumber\\
 &\left.\Big( \chi_{\mathbf{k}}(\tau)\beta(\mathbf{k}) +\chi^*_{\mathbf{k}}(\tau)\beta^\dag(\mathbf{-k}) \Big)
 \Big( \chi_{\mathbf{q}}(\tau)\beta(\mathbf{q}) +\chi^*_{\mathbf{q}}(\tau)\beta^\dag(\mathbf{-q}) \Big)
 \right\}.
\end{align}
In order to compute the two-point correlation function $\langle\chi(\mathbf{x},t)\chi(\mathbf{y},t)\rangle$ of the inflaton, we use the formula 
%introduced in \cite{Weinberg}
\begin{equation}\label{calka}
\langle\chi(\mathbf{x},t)\chi(\mathbf{y},t)\rangle\simeq\langle\chi_I(\mathbf{x},t)\chi_I(\mathbf{y},t)\rangle+
i\int^t_{-\infty}dt'e^{-\epsilon'|t'|}\langle[\mathcal{H}_\text{int}(t'),\chi_I(\mathbf{x},t)\chi_I(\mathbf{y},t)] \rangle,
\end{equation}
where $\epsilon'$ is an infinitesimal parameter that cuts off the early time part of the integration. The relation (\ref{calka}) was originally introduced in
\cite{Weinberg}, the so-called $\langle in - in \rangle$ formalism applied to the calculation of higher order Gaussian and non-Gaussian corrections in cosmology.

To commence with, we calculate 
the commutator appearing in the relation (\ref{calka}). Namely, one obtains
\begin{equation}
 [\mathcal{H}_\text{int}(t'),\chi_I(\mathbf{x},t)\chi_I(\mathbf{y},t)]=[\mathcal{H}_1(t'),\chi_I(\mathbf{x},t)\chi_I(\mathbf{y},t)]+
 [\mathcal{H}_2(t'),\chi_I(\mathbf{x},t)\chi_I(\mathbf{y},t)],
\end{equation}
where we have denoted by $\mathcal{H}_1(t')$ the part including $\dot{\chi}$, while by $\mathcal{H}_2(t')$ the rest of (\ref{int}). It implies
\begin{align}\label{wyn1}
 &[\mathcal{H}_1(t'),\chi_I(\mathbf{x},t)\chi_I(\mathbf{y},t)]\nonumber\\&=
 \frac{3}{2}bH\int\frac{d^3x'}{\tau'}\int \frac{d^3k}{(2\pi)^3}\int\frac{d^3q}{(2\pi)^3}e^{i\mathbf{x}'(\mathbf{k}+\mathbf{q})
 -i(\mathbf{kx}+\mathbf{qy})}\left((\tau+\frac{i}{q})(\tau+\frac{i}{k})e^{-i(\mathbf{q+k})(\tau'-\tau)}-\text{h.c.} 
 \right),
\end{align}
and for the $\mathcal{H}_2(t')$-part, one deals with the following relation:
\begin{align} \label{above}
 &[\mathcal{H}_2(t'),\chi_I(\mathbf{x},t)\chi_I(\mathbf{y},t)] \\&=
 \frac{bH^3}{2}\int\frac{d^3x'}{\tau'}\int \frac{d^3k}{(2\pi)^3}\int\frac{d^3q}{(2\pi)^3}\frac{e^{i\mathbf{x}'(\mathbf{k}+\mathbf{q})
 -i(\mathbf{kx}+\mathbf{qy})}}{kq}\Big[ \frac{(k_xy-k_yx)(q_xy-q_yx)}{x^2+y^2}+k_zq_z\nonumber\\
 &+k_\perp q_\perp \cos(\theta-\theta_k)\cos(\theta-\theta_q)
 \Big]\left((\tau'-\frac{i}{k})(\tau'-\frac{i}{q})
 (\tau+\frac{i}{k})(\tau+\frac{i}{q})e^{-i(\mathbf{q+k})(\tau'-\tau)}-\text{h.c.} 
 \right), \nonumber
\end{align}
where one decomposes the vectors along $z$-axis and the two-dimensional subspace, perpendicular to it.

Let us notice that the first term in the square bracket in the equation (\ref{above}) has been already calculated in \cite{string}. The rest of 
the formula in question, after 
the change of the time coordinate $dt'=-\frac{d\tau'}{H\tau'}$ on the conformal time, and integrating over $\tau$, yields
\begin{align}\label{wyn2}
 \langle\chi(\mathbf{x},t)\chi(\mathbf{y},t)\rangle_{2_\text{part}}
 &=-bH^2\int d^3x'\int \frac{d^3k}{(2\pi)^3}\int\frac{d^3q}{(2\pi)^3}e^{i\mathbf{x}'(\mathbf{k}+\mathbf{q})
 -i(\mathbf{kx}+\mathbf{qy})}\Big[k_zq_z
 +\frac{(\mathbf{k}_\perp\mathbf{x}'_\perp)(\mathbf{q}_\perp\mathbf{x}'_\perp)}{\mathbf{x}'^{2}_\perp}
 \Big]\nonumber\\
 &\times\left(\frac{kq+q^2+k^2+\tau k^2q^2}{k^3q^3(k+q)}
 \right).
\end{align}
On the other hand, 
the result of the computations of the first term is given in \cite{string}, by the following:
\begin{align}
 \langle\chi(\mathbf{x},t)\chi(\mathbf{y},t)\rangle_{2_\text{part2}}
 &=-bH^2\int d^3x'\int \frac{d^3k}{(2\pi)^3}\int\frac{d^3q}{(2\pi)^3}e^{i\mathbf{k}(\mathbf{x}'-\mathbf{x})
 +i\mathbf{q}(\mathbf{x}'-\mathbf{y})}\nonumber\\
 &\times\left(\frac{kq+q^2+k^2+\tau k^2q^2}{k^3q^3(k+q)}
 \right)\left(\frac{(y'k_x-x'k_y)(y'q_x-x'q_y)}{x'^2+y'^2}\right).
\end{align}

Similarly,  for  the relation (\ref{wyn1}), one obtains
\begin{align}\label{wyn1a}
  \langle\chi(\mathbf{x},t)\chi(\mathbf{y},t)\rangle_{1}=2bH
  \int d^3x'\int \frac{d^3k}{(2\pi)^3}\int\frac{d^3q}{(2\pi)^3}e^{i\mathbf{x}'(\mathbf{k}+\mathbf{q})-i(\mathbf{kx}+\mathbf{qy})}
  \frac{kq\tau^2-1}{kq(k+q)}.
\end{align}
Taking into account the result presented in \cite{string} and ours given by the equations  (\ref{wyn2}) and (\ref{wyn1a}), finally we arrive at the formula
\begin{align}
 \Delta\langle\chi(\mathbf{x},t)\chi(\mathbf{y},t)\rangle
 &=b\int d^3x'\int \frac{d^3k}{(2\pi)^3}\int\frac{d^3q}{(2\pi)^3}e^{i\mathbf{x}'(\mathbf{k}+\mathbf{q})-i(\mathbf{kx}+\mathbf{qy})}
 \left(3H\frac{kq\tau^2-1}{kq(k+q)}\right.\nonumber\\
 &\left.-H^2\mathbf{k} \cdot\mathbf{q}\Big( \frac{kq+q^2+k^2+k^2q^2\tau^2}{k^3q^3(k+q)} \Big)
 \right).
\end{align}
After integrating over $x'$, $ \Delta\langle\chi(\mathbf{x},t)\chi(\mathbf{y},t)\rangle$ is provided by
\begin{align}\label{chi}
  \Delta\langle\chi(\mathbf{x},t)\chi(\mathbf{y},t)\rangle
 &=3Hb\int \frac{d^3k}{(2\pi)^3}\int\frac{d^3q}{(2\pi)^3}e^{-i(\mathbf{kx}+\mathbf{qy})}(2\pi)^\frac{3}{2}\delta(\mathbf{k+q})
 \frac{kq\tau^2-1}{kq(k+q)}\\
 &-bH^2\int \frac{d^3k}{(2\pi)^3}\int\frac{d^3q}{(2\pi)^3}e^{-i(\mathbf{kx}+\mathbf{qy})}(2\pi)^\frac{3}{2}
 \Big[ \frac{kq+q^2+k^2+k^2q^2\tau^2}{k^3q^3(k+q)} \Big]\mathbf{k\cdot q}. \nonumber
\end{align}
We notice that the first term is translational invariant because of the presence of the Dirac delta function while rotational invariance is broken since the power 
spectrum depends on both vector fields magnitudes $k$ and $q$. The second term violates the translational invariance by the presence of 
the special point $\mathbf{x}_0$ (which is taken at the origin) and preserves the rotational 
invariance about this point \cite{carrol}.
Thus, if we approximate the density perturbations as $\delta=\kappa\chi$, the part of the power spectrum violating rotational and translational invariance,
which might be visible now ($\tau=0$) is written as 
\begin{equation}\label{result}
 \Delta P(k,q,\mathbf{k\cdot q})= -3(2\pi)^\frac{3}{2}b\kappa^2H\left(\frac{\delta(\mathbf{k+q})}{kq(k+q)} +\frac{1}{3}
 \Big[ \frac{kq+q^2+k^2}{k^3q^3(k+q)} \Big]\mathbf{k\cdot q}
 \right).
\end{equation}
The relation (\ref{result}) constitutes the main result of our considerations. It will be discussed in what follows.

%%%%%%%%%%%%%%%%%%%%%%%%%%%%%%%%%%%%%%%%%%%%%%%%%%%%%%%%%%%%%%%%%%%%%%%%%%%%%%%%%%%%%%%%%%%%%%%%%%%%%%%%%%%%%%%%
\section{Discussion and conclusions}\label{sec3}
We have computed the influence on the energy density perturbations of our Universe which is caused by the simple wormhole (\ref{metric}), with 
the shape parameter 
$b(r)=b=\text{constant}$ and 
$m=1$,that could exist during the inflation epoch and have passed our horizon volume. Calculating of the two-point correlation function of the free massless scalar field (\ref{calka})
revealed that the effect of the wormhole on the perturbations of the energy density envisages in violating translational and rotational invariance. 
It leads to the conclusion that the obtained result (\ref{result})
may be potentially  found in the anisotropy data of CMB.

\begin{figure}[h!]
	\centering
	\includegraphics[width=0.80\textwidth]{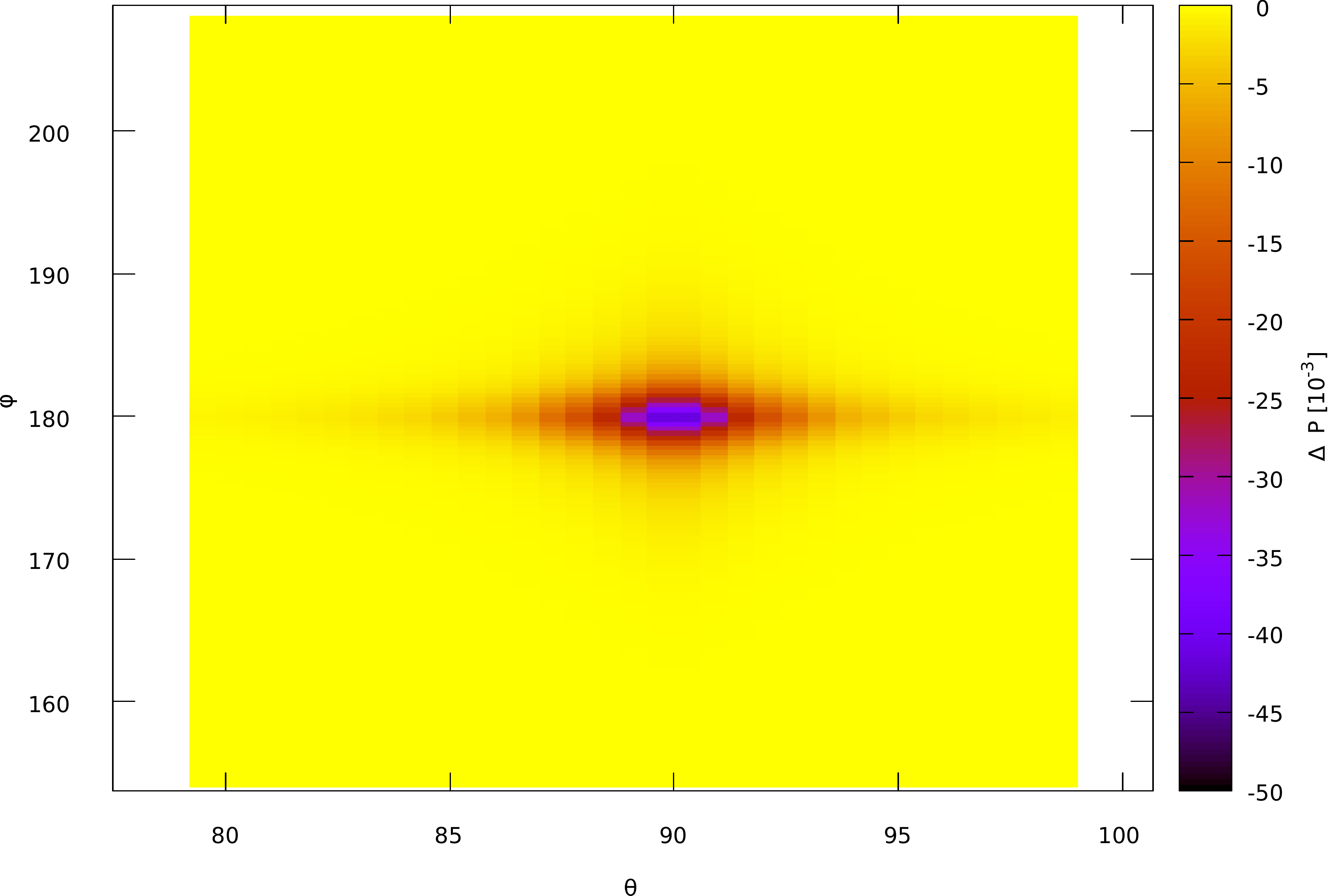}
	\caption{
	\label{fig.modecenter}
	Map of $\Delta P(k,q,\mathbf{k\cdot q})$ showing the value of the function near initial peak ($k = q = 0.1$) with respect to different
	orientations of $\mathbf{q}$ described in terms of $(\theta, \phi)$ angles. The lowest value of power spectrum perturbation occurs 
	for $\theta = 90^{\circ}, \phi = 180^{\circ}$ which corresponds to anti-parallel $\mathbf{k} = -\mathbf{q}$ setting.}
\end{figure}

Let us analyze the behavior of the fluctuation of the power spectrum. The formula (\ref{result}) in the general case depends on the mode wave 
vectors $(\mathbf{k,q})$ which for convenience we express in Hubble parameter units. Let us notice that if the inflation duration is assumed to last about $60$
e-folds, the wavelength of the modes with $k/H=1$ is about the size of the Universe.
The dependence on the conformal time is suppressed in (\ref{result}), because of $\tau=0$ in the current evolutionary phase, but $\tau$-dependence can be easily obtained from 
(\ref{chi}).
In all the plots presented here we took $\kappa = 10^{-3}$ and $b$ as unity, moreover the Dirac delta function will be approximated by the Lorentzian $\delta ( k ) \simeq \zeta / [ \pi ( k^{2} + \zeta^{2} ) ]$ with 
broadening $\zeta = 0.0025$. The influence of the configuration of $\mathbf{(k,q)}$ modes on the function is as expected, that is, 
we confirm for $\tau = 0$ that $\Delta P$ is large for small values of $k$ and $q$ due to the term with Dirac delta function
and it decreases rapidly with growth of the arguments. Therefore, the size of the initial peak is strongly dependent on a
spatial configuration of the vectors which is present in the scalar product.
To investigate it, without loss of generality, we set $\mathbf{k}$ to be oriented along the $x$-axes direction in the Fourier space, then we 
describe $\mathbf{q}$ vector with spherical coordinate angles $(\theta, \phi)$. By rotating $\mathbf{q}$ we plot a map of values dependent
on the aforementioned angles (figure \ref{fig.modecenter}).

%%%%%%%%%%%%%%%%%%%%%%%%%%%%%%%%%%%%%%%%%%%%%%%%%%%%%%%%%%%%%%%%%%%
We find out that the biggest perturbations occur for $\mathbf{k} = -\mathbf{q}$ setting which confirms the statement
from \cite{carrol}. Therefore, we focus on this configuration and we plot the values of $\Delta P$ with increasing value of $k$. 

\begin{figure}[h]
	\centering
	\includegraphics[width=0.80\textwidth]{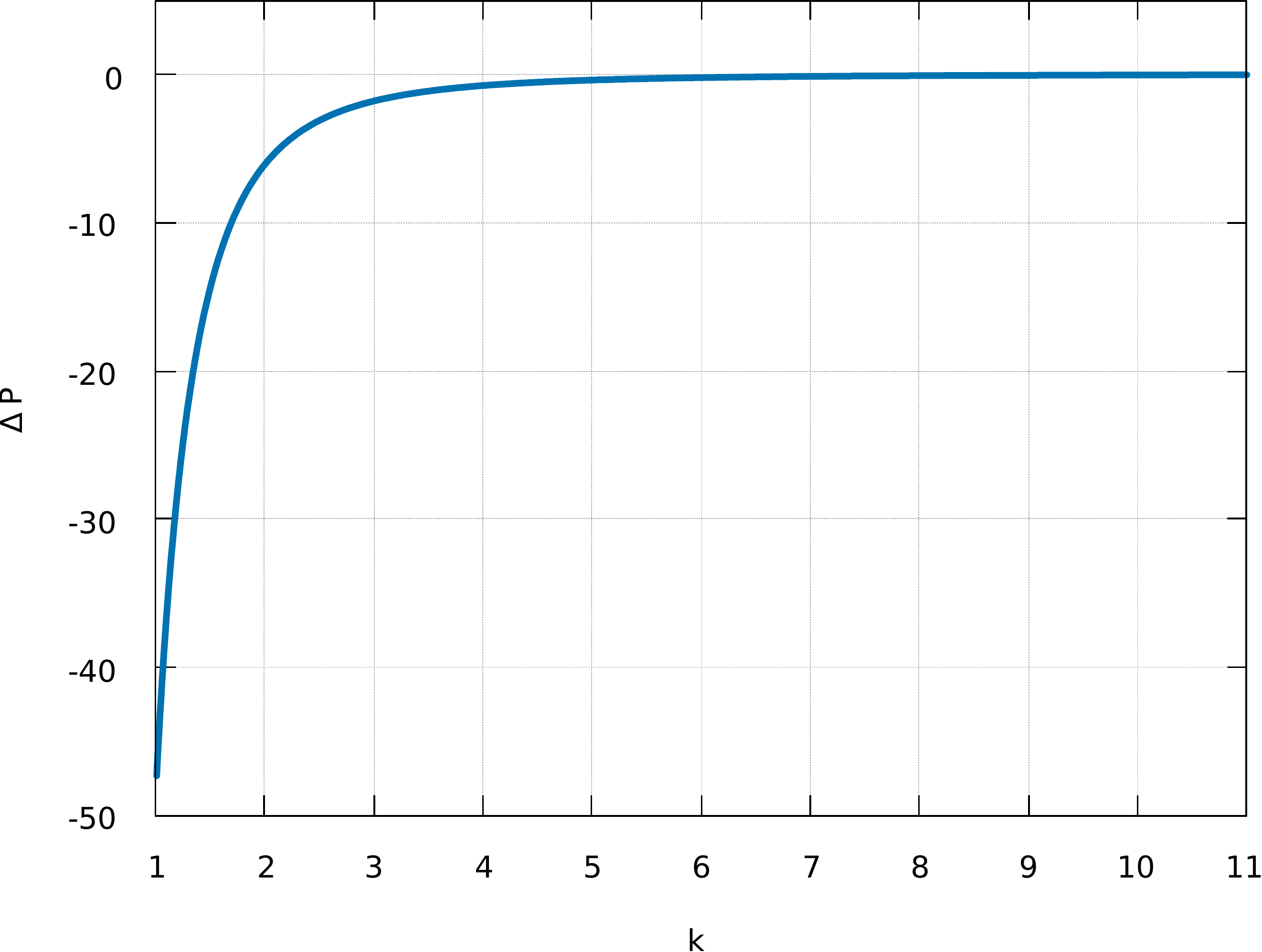}
	\caption{
	\label{fig.k_t=0}
	$\Delta P(k,q,\mathbf{k\cdot q})$ with $\tau = 0$ for different values of $k$. Function rapidly vanishes with 
	increasing argument. Only the modes with small values of $k$ can have an influence on the anisotropy spectrum 
	observed nowadays ($\tau = 0$).}
\end{figure}

In figure  \ref{fig.k_t=0} we plot the power spectrum function for $\tau = 0$ versus the different values of the momenta $\bf k$. It can be seen that for the
increasingly values of $\bf k$, the function $\Delta P$ rapidly tends to its zero value. The only modes responsible for small values of the momenta,
influence the anisotropy spectrum observed nowadays.

\begin{figure}[h]
	\centering
	\includegraphics[width=0.80\textwidth]{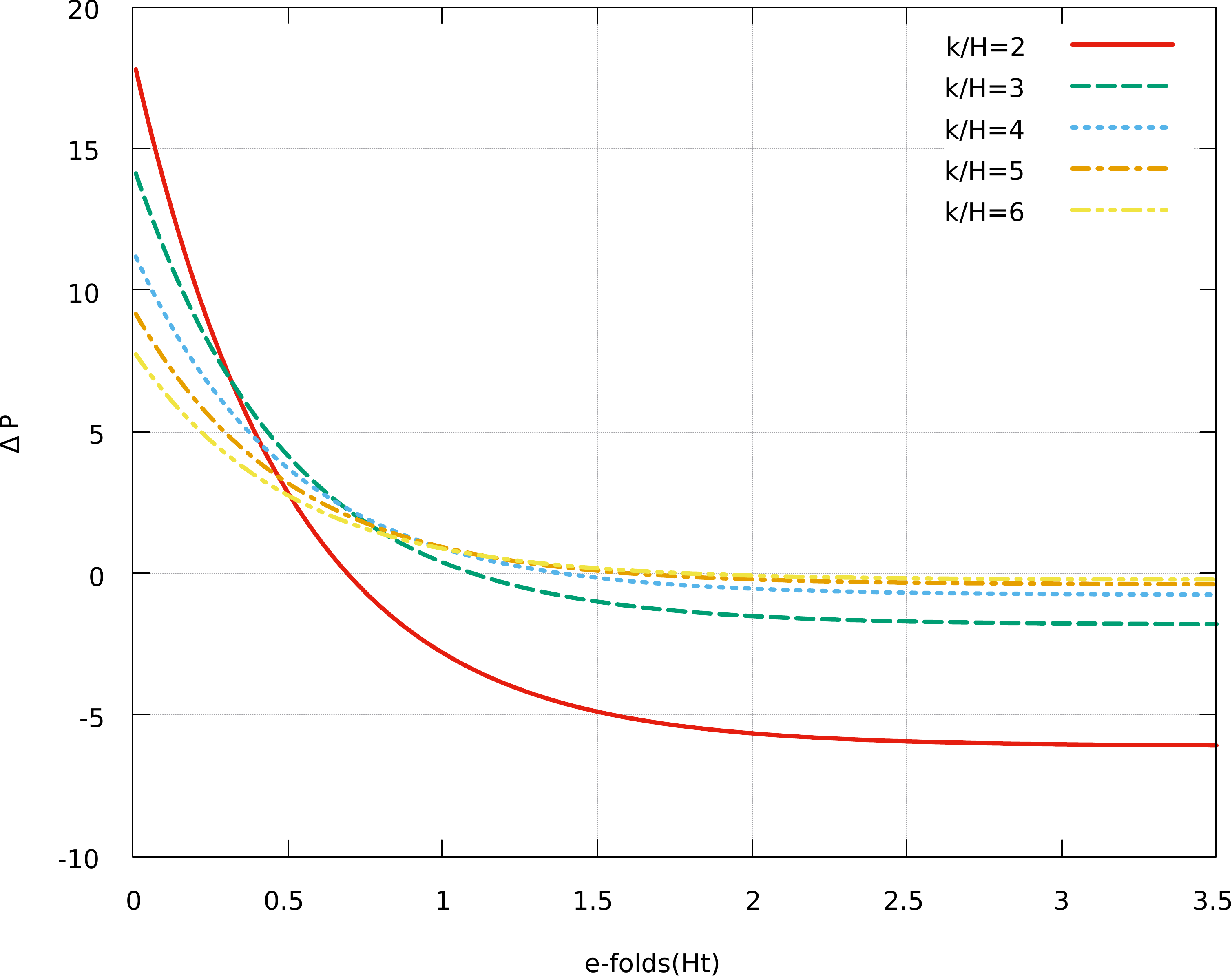}
	\caption{
	\label{fig.dP_t}
	Power spectrum perturbation versus e-folding number connected with the conformal time $\tau = -\frac{e^{-Ht}}{H}$. Initial fluctuations decrease to their final values
	after approximately three e-folds.}
\end{figure}

Finally, in figure \ref{fig.dP_t},  we plot the time dependence of power spectrum perturbation for a few different e-folds. Consequently
we draw a conclusion that the longer inflation lasts, the weaker effects of the studied wormhole we observe. It can be envisaged that
$\Delta P$-function rapidly decreases to its final value after about three e-folds.

One concludes that the observed CMB anisotropies deliver a new research window onto physics and phenomena that may occur during the rapid expansion phase of the
early Universe. The resulting theoretical claims can be compared with the observations, giving us the new opportunity to test
our models of  the early history of the Universe.

%%%%%%%%%%%%%%%%%%%%%%%%%%%%%%%%%%%%%%%%%%%%%%%%%%%%%%%%%%%%%%%%%%%%%%%%%%%%%%%%%%%%%%%%%%%%%%%%%%%%%%%%%%%%%%%%

%%%%%%%%%%%%%%%%%%%%%%%%%%%%%%%%%%%%%%%%%%%%%%%%%%%%%%%%%%%%%%%%%%%%%%%%%%%%%%

%\section{Sections}
%\subsection{And subsequent}
%\subsubsection{Sub-sections}
%\paragraph{Up to paragraphs.} We find that having more levels usually
%reduces the clarity of the article. Also, we strongly discourage the
%use of non-numbered sections (e.g.~\texttt{\textbackslash
%  subsubsection*}).  Please also see the use of
%``\texttt{\textbackslash texorpdfstring\{\}\{\}}'' to avoid warnings
%from the hyperref package when you have math in the section titles

%\appendix
%\section{Some title}
%Please always give a title also for appendices.

%%%%%%%%%%%%%%%%%%%%%%%%%%%%%%%%%%%%%%%%%%%%%%%%%%%%%%%%%%%%%%%%%%%%%%%%%%%%%%%%%%%%%%%%%%%%%%%%%%%%%%%%%%%%%%%%%%%%%%%%%%%%%%%%%%%%%%%%%%%%%%%%%%%%%%%%%%%%

\acknowledgments
MR and AW were partially supported by the grant of the National Science Center \\
$DEC-2014/15/B/ST2/00089$. AW is also partially supported by FAPES (Brazil).

%%%%%%%%%%%%%%%%%%%%%%%%%%%%%%%%%%%%%%%%%%%%%%%%%%%%%%%%%%%%%%%%%%%%%%%%%%%%%%
%\paragraph{Note added.} This is also a good position for notes added
%after the paper has been written.


%%%%%%%%%%%%%%%%%%%%%%%%%%%%%%%%%%%%%%%%%%%%%%%%%%%%%%%%%%%%%%%%%%%%%%%%%%%%%%%
\end{document}